\documentclass[manuscript]{aastex}

\slugcomment{Not to appear in Nonlearned J., 45.}

\shorttitle{Collapsed Cores in Globular Clusters}
\shortauthors{Djorgovski et al.}

\begin{document}


\title{ $\Lambda$CDM\,, $\Lambda$DGP and extended phantom-like cosmologies}


\author{Kourosh Nozari \altaffilmark{1}}
\affil{Isalamic Azad University, Sari Branch, Sari, IRAN} \and
\author{Faeze Kiani \altaffilmark{2}}
\affil{Department of Physics,
Faculty of Basic Sciences,\\
University of Mazandaran,\\
P. O. Box 47416-95447, Babolsar, IRAN}


\altaffiltext{1}{knozari@umz.ac.ir}
\altaffiltext{2}{fkiani@umz.ac.ir}


\begin{abstract}
In this paper we compare outcomes of some extended phantom-like
cosmologies with each other and also with $\Lambda$CDM\, and
$\Lambda$DGP. We focus on the variation of the luminosity distances,
the age of the universe and the deceleration parameter versus the
redshift in these scenarios. In a dynamical system approach, we show
that the accelerating phase of the universe in the $f(R)$-DGP
scenario is stable if one consider the \emph{curvature fluid} as a
phantom scalar field in the equivalent scalar-tensor theory,
otherwise it is a transient and unstable phenomenon. Up to the
parameters values adopted in this paper, the extended
$F(R,\varphi)$-DGP scenario is closer to the $\Lambda$CDM scenario
than other proposed models. All of these scenarios explain the
late-time cosmic speed-up in their normal DGP branches, but the
redshift at which transition to the accelerating phase occurs are
different: while the $\Lambda$DGP model transits to the accelerating
phase much earlier, the $F(R,\varphi)$-DGP model transits to this
phase much later than other scenarios. Also, within the parameter
spaces adopted in this paper, the age of the universe in the
$f(R)$-DGP model is larger than $\Lambda$CDM, but this age in
$F(G,\varphi)$-DGP is smaller than $\Lambda$CDM.\\
\end{abstract}


\keywords{Braneworld Cosmology, Phantom Mimicry, Dynamical System}



\section{Introduction}

One of the most remarkable discoveries of the past decade is that
the universe is currently in an accelerating phase, which means that
expanding velocity of the universe is increasing. This phenomenon is
supported by data related to the luminosity measurements of high red
shift supernovae  \citep{Per99,Rie98,Ast06,Woo07}, measurement of
degree-scale anisotropies in the cosmic microwave background (CMB)
\citep{Mil99,Han00,Spe03} and large scale structure (LSS)
\citep{Col01,Teg04,Col05,Spr06}. The rigorous treatment of this
phenomenon can be provided in the framework of general relativity.
In the expression of general relativity, late time acceleration can
be explained either by an exotic fluid with large negative pressure
that is dubbed as \textit{dark energy} in literature, or by
modifying the gravity itself which is dubbed as \textit{dark
geometry} or \textit{dark gravity} proposal. The first and simplest
candidate of dark energy is the cosmological constant, $\Lambda$
\citep{Sah00,Pad03,Cop06}. But there are theoretical problems
associated with it, such as its unusual small numerical value ( the
fine tuning problem), no dynamical behavior and its unknown origin
\citep{Wei89,Car01,Cal99}. These problems have forced cosmologists
to introduce alternatives in which dark energy evolves during the
universe evolution. Scalar field models with their specific features
provide an interesting alternative for cosmological constant and can
reduce the fine tuning and coincidence problems. In this respect,
several candidate models have been proposed: "quintessence" scalar
field \citep{Rat88,Sai00,Bra00,Bar00,Sah00,Sah02,Sam03}, phantom
fields \citep{Cal02,Tsu04,Cal05, Cai10, Moy09} and chaplygin gas
\citep{Kam01,Dev03,Ame03,Roo07,Bou07,Zha06a,Zha06b,Ber04,Bie05,Hey08,Roo08a,Roo08b,Zha09,Set09}
are among these candidates.

As an alternative for dark energy, modification of gravity can be
accounted for the late time acceleration. Among the most popular
modified gravity scenarios which may successfully describe the
cosmic speed-up, is $f(R)$ gravity
\citep{Cap03,Sot10,Noj03,Noj04,Noj05,Noj06,Noj07,Noj08,Bam08,Car04,Ame07,Noz08a,Ata08,Saa09,
Set08}. Modified gravity also can be achieved by extra-dimensional
theory in which the observable universe is a 4-dimensional brane
embedded in a five-dimensional bulk. Dvali- Gabadadze-Porrati (DGP)
model is one of the extra-dimensional models that can describe
late-time acceleration of the universe in its self-accelerating
branch due to leakage of gravity to the extra dimension
\citep{Dva00a,Dva00b,Dva01,Dva02,Def01,Lue06}.

Recent observations constrain the equation of state parameter of the
dark energy to be $\omega_{X}\approx-1$ and even $\omega_{X}<-1$
\citep{Mel03,Rie04,Kom09}. One of the candidates for dark energy of
this kind is the phantom scalar field. This component has the
capability to create the mentioned acceleration and its behavior is
extremely fitted to observations. But it suffers from problems; it
violates the null energy condition and its energy density increases
with expansion of the universe which is an unphysical behavior. Also
it causes the quantum vacuum instabilities. So, cosmologists have
tried to realize a kind of phantom-like behavior in the cosmological
models without introduction of phantom fields
\citep{Sah03,Sah04,Sht00,Lue04}. With phantom mimicry ( the
phantom-like behavior), effective equation of state parameter of
dark component remains less than $-1$ and effective energy density
of the unverse increases with cosmic expansion. In the framework of
general relativity these two expressions are equivalent. However, in
the effective picture of cosmological models, in order to satisfy
phantom-like behavior, the mentioned two expressions must be
investigated separately. The phantom mimicry discussed in this study
has a geometric origin. In this paper, we discuss briefly
DGP-inspired theoretical models that realize phantom-like behavior.
We present cosmological dynamics in each proposed model and then we
compare these models with $\Lambda$CDM and also with each other
through investigation of their expansion histories. Within a
dynamical system ( phase space) approach, we show that the
accelerating phase of the universe in the $f(R)$-DGP scenario is
stable if one consider the curvature fluid as a phantom scalar field
in the equivalent scalar-tensor theory, otherwise it is a transient
and unstable phenomenon. As another important probe, we study the
age of the universe in each model. We show that some of these models
account for a transient accelerated phase.

\section{The phantom-like evolution}


As we have pointed out in the introduction, the observational
evidences show that the equation of state parameter of dark energy
in the universe can be less than $-1$. Many attempts have been made
to find dark energy models that allow the so-called phantom dark
energy: a dark energy component with large negative pressure and
negative kinetic energy with equation of state parameter less than
$-1$. A phantom field is described by the following action
\begin{equation}
S=\int \Big[\frac{1}{2}g^{\mu\nu}\partial_{\mu}\varphi\partial_{\nu}
\varphi-V(\varphi)\Big]\sqrt{-g}d^{4}x\,,
\end{equation}
The kinetic energy term of the phantom field in the corresponding
lagrangian enters with opposite sign in contrast to the ordinary
matter and this distinguishes phantom field from ordinary
(canonical) fields. But this distinctive property of the phantom
field causes a series of quantum vacuum instabilities. Also, due to
characteristic $p<-\rho$ feature of this kind of matter, it violates
the null energy condition. On the other hand, dark energy with
$\omega<-1$ (phantom energy) is beset with a host of undesirable
properties which makes this model of dark energy to be unattractive.
So, cosmologists attempted to find a way which removes these
problems. They are forced to consider models that realize
phantom-like behavior without introducing any phantom matter. In the
language of general relativity, phantom-like behavior means that the
equation of state parameter of dark energy is less than $-1$ or its
energy density increases with expansion of the universe. In the
cosmological models that there is no phantom matter, two mentioned
expressions for the phantom-like behavior are not equivalent and
should be investigated separately.




\section{Phantom mimicry: extended models}

\subsection{$\Lambda$CDM vs $\Lambda$DGP} \label{bozomath}


A cosmological model which has the capability to realize the
phantom-like behavior without introducing any phantom matter, is the
$\Lambda$DGP model. Firstly, we introduce briefly the $\Lambda$DGP
model. The DGP model, as a modified theory of gravity that modifies
the geometric sector of the Einstein field equations, has a modified
Friedmann equation as follows \citep{Dva00a,Def01,Lue06}
\begin{equation}
H^{2}\pm\frac{H}{r_{c}}=\frac{8\pi G}{3}\rho(t)\,,
\end{equation}
where $H(t)=\frac{\dot{a}}{a}$ is the Hubble parameter. This model,
in its self-accelerating branch (corresponding to the negative sign
of equation (2)), explains late-time accelerated expansion of the
universe. But, the normal branch (positive sign), has no
self-accelerating behavior. Nevertheless, the self-accelerating
branch has beset with ghost problem. The most trivial extended-DGP
model which can create the accelerated phase in the normal branch is
the $\Lambda$DGP model. In this model a cosmological constant which
plays the role of dark energy, lies on the brane. This scenario
based on the so-called \textit{dynamical screening} of the brane
cosmological constant, has the capacity to realize a phantom-like
behavior without introducing any phantom matter neither on the brane
nor in the bulk. In this case, brane cosmological constant as a dark
energy component is screened due to curvature modification at
late-time \citep{Sah03,Sah04,Sht00,Lue04}. The action of this model
is
\begin{equation}
{\cal{S}}=\frac{1}{2} M_{5}^{3} \int d^{5}x\sqrt{-g}{\cal{R}}+
\frac{1}{2}m_{p}^{2}\int d^{4}x\sqrt{-q}R+\int
d^{4}x\sqrt{-q}{\cal{L}}_{m}\,,
\end{equation}
where $M_{5}$ is the five-dimensional fundamental scale. The first
term in ${\cal{S}}$ is the Einstein-Hilbert action in five
dimensions for a five-dimensional metric $g_{AB}$ with Ricci scalar
${\cal{R}}$. The metric $q_{\mu\nu}$ is the induced
(four-dimensional) metric on the brane, and $q$ is its determinant.
The second integral contains an induced curvature effect that
appears in DGP setup due to quantum corrections via interaction of
the bulk graviton with matter on the brane. To apply correct
boundary conditions, the Gibbons-Hawking term containing the
extrinsic curvature of the brane should be added to the brane part
of the action. The crossover distance, that gravity in the scales
larger than it appears to be $5$-dimensional, is defined as follows
\begin{equation}
r_{c}=\frac{m_{p}^{2}}{2M_{5}^{3}}
\end{equation}
The cosmology on the brane for a spatially-flat universe follows
Deffayet's modified Friedmann equation \citep{Dva00a,Def01,Lue06}
\begin{equation}
H^{2}\pm \frac{H}{r_{c}}=\frac{8\pi
G}{3}(\rho_{m}+\rho_{\Lambda})\,.
\end{equation}
The negative and positive signs in this equation represent the
self-accelerating and normal branches respectively. In which
follows, we focus on the normal branch, because this branch has the
key property that brane is extrinsically curved so that shortcuts
through the bulk allow gravity to screen the effect of the brane
energy-momentum contents at Hubble parameter $H\sim r_{c}^{-1}$ and
there is no ghost instability in this branch
\citep{Sah03,Sah04,Sht00,Lue04}
\begin{equation}
H^{2}+\frac{H}{r_{c}}=\frac{8\pi
G}{3}\Big(\rho_{m}+\rho_{\Lambda}\Big).
\end{equation}
Comparing this equation and the standard Friedmann equation in a
spatially flat universe
\begin{equation}
H^{2}=\frac{8\pi G}{3}(\rho_{m}+\rho_{DE})\,,
\end{equation}
one can find an effective dark energy component as follow
\citep{Mel03,Rie04,Kom09}
\begin{equation}
\frac{8\pi G}{3}\rho_{DE}^{eff}=\Lambda-\frac{H}{r_{c}}.
\end{equation}
Since in this case $H(t)$ is a decreasing function of the cosmic
time, the effective dark energy component increases with time, and
therefore we realize a phantom-like behavior without introducing any
phantom matter that violets the null energy condition and suffers
from several theoretical problems. We can define a $w_{eff}$ by
using $\rho_{eff}=\rho_{0}(1+z)^{3(1+w_{eff})}$  and comprehend that
in a red-shift such as $z^{*}$ which $H(z^{*})<\Lambda r_{c}$ (
during the epoch which dark energy is dominated ), $w_{eff}$ is less
than $-1$ \citep{Laz06,Maa06,Ala06,Laz07}. In other words, in the
mentioned red-shift, two expressions of the phantom-like behavior
are equivalently applicable. Also, existence of $\frac{H}{r_{c}}$
term causes the cosmological constant $\Lambda$ to be screened, and
to be appeared less than its actual value. This case is called the
\emph{gravitational screening effect} and as we have pointed out,
one can realize phantom-like behavior based on this effect
\citep{Sah03,Sah04,Sht00,Lue04}. The $\Lambda$DGP model is an
alternative for $\Lambda CDM$ ( the cosmological constant $\Lambda$
plus the cold dark matter ). It has been shown recently that
$\Lambda$DGP in some respects gives even better fit to observations
than other dark energy scenarios, albeit with one more parameter
\citep{Sol09}. To investigate expansion history of $\Lambda$DGP and
comparing it with constant-$w$ dark energy models (such as the
$\Lambda$CDM), we study variation of luminosity distances versus the
redshift in these scenarios. By rewriting  the Friedmann equation
(6) in the more phenomenological form, we find
\begin{equation}
\frac{H(z)}{H_{0}}=\sqrt{\Omega_{m}(1+z)^{3}+\Omega_{\Lambda}+\Omega_{r_{c}}}-\sqrt{\Omega_{r_{c}}}
\end{equation}
Where
\begin{equation}
\Omega_{m}=\frac{8\pi G}{3H_{0}^{2}}\rho_{0m},\ \quad
\Omega_{\Lambda}=\frac{\Lambda}{3H_{0}^{2}},\ \quad
\Omega_{r_{c}}=\frac{1}{4r_{c}^{2}H_{0}^{2}}\,.
\end{equation}
Also, by using this dimensionless equation, one can imply a
constrain on the parameters of the model as follows
\begin{equation}
\Omega_{m}+\Omega_{\Lambda}=1+2\sqrt{\Omega_{r_{c}}}\,.
\end{equation}
This constraint means that a flat $\Lambda DGP$ model mimics a
closed $\Lambda CDM$ model in the $(\Omega_{m},\Omega_{\Lambda})$
plane \citep{Laz06,Maa06,Ala06,Laz07}. Friedmann equation in the
$\Lambda CDM$ model ( constant-$w$ model with $w=-1$ ) is
phenomenologically written as follows
\begin{equation}
\frac{H(z)}{H_{0}}=\sqrt{\Omega_{m}(1+z)^{3}+(1-\Omega_{m})(1+z)^{3(1+w)}}.
\end{equation}
By considering luminosity-distance in the standard form in spatially
flat cosmology
\begin{equation}
d_{L}(z)=(1+z)\int_{0}^{z} \frac{H_{0}}{H(z)}dz \,,
\end{equation}
we can compare $d_{L}(z)$ calculated in $\Lambda$DGP and
$\Lambda$CDM frameworks. This comparison is shown in figure $3$. Up
to parameters values used to plot this figure, the luminosity
distance calculated in the $\Lambda DGP$ is larger than that of the
$\Lambda CDM$ for a given redshift.




\section{$f(R)$-DGP scenario}


As an extension of the previous subsection, here we consider
possible modification of the induced gravity on the brane in the
spirit of $f(R)$ theories
\citep{Noz08a,Ata08,Saa09,Noz09a,Noz09b,Noz09d,Noz09c,Bou09,Ata06,Ata07,Ata09}.
We assume that induced gravity on the brane can be modified by a
general $f(R)$ term. It has been shown that $4D$ $f(R)$ theories can
follow closely the expansion history of the $\Lambda$CDM universe
\citep{Hu07,Mar09}. Here we study an extension of $f(R)$ theories to
a DGP braneworld setup. By focusing on the luminosity
distance-redshift relation, we compare expansion history of this
type of model with other alternative scenarios. The action of this
model can be written as follows
\begin{equation}
{\cal{S}}=\frac{M_{5}^{3}}{2}\int d^{5}x\sqrt{-g} {\cal{R}} + \int
d^{4}x\sqrt{-q}\Big(M_{5}^{3}\overline{K}+{\cal{L}}\Big)\,,
\end{equation}
where by definition
\begin{equation}
{\cal{L}}=\frac{m_{p}^{2}}{2}f(R)+{\cal{L}}_{m}\,.
\end{equation}
By calculating the bulk-brane Einstein's equations and using a
spatially flat FRW line element, the following modified Friedmann
equation is obtained
\citep{Noz09a,Noz09b,Noz09c,Noz09d,Ata06,Ata07,Ata09,Bou09}
\begin{equation}
H^{2}=\frac{8\pi
G}{3}(\rho^{m}+\rho^{(curv)})\pm\frac{H}{\bar{r}_{c}}
\end{equation}
where
\begin{equation}
\rho^{(curv)}=m_{p}^{2}\bigg(\frac{1}{2} \Big[f(R)-R f'(R)\Big]
-3\dot{R}H f''(R) \bigg)\,,
\end{equation}
is energy density corresponding to the curvature part of the theory.
This energy density can be dubbed as \emph{dark curvature} energy
density. $\bar{r}_{c}$ is the re-scaled crossover distance that is
defined as $\bar{r}_{c}=r_{c}f'(R)$ and a prime marks
differentiation with respect to the Ricci scalar, $R$. We note that
in this scenario there is an effective gravitational constant, which
is re-scaled by $f'(R)$ so that $G=G_{eff}\equiv\frac{1}{8\pi
m_{p}^{2}f'(R)}$ \citep{Noz09a}. In order to compare this model with
other alternative scenarios, it is more suitable to rewrite the
normal branch of the Friedmann equation (16) in the following form
\begin{equation}
\frac{H(z)}{H_{0}}=\sqrt{\Omega_{m}(1+z)^3+
\Omega_{curv}(1+z)^{3(1+\omega_{curv})}+\Omega_{r_{c}}}-\sqrt{\Omega_{r_{c}}}\,,
\end{equation}
where by definition $$\Omega_{curv}=\frac{8\pi
G}{3H_{0}^{2}}\rho_{0}^{(curv)}\,\,,\quad\quad
\Omega_{r_{c}}=\frac{1}{4[r_{c}f_{0}'(R)]^{2}H_{0}^{2}}\,,$$ and
also

\begin{equation}
w_{curv}=-1+\frac{\ddot{R}f''(R)+\dot{R}\Big[\dot{R}f'''(R)-
Hf''(R)\Big]}{\frac{1}{2}[f(R)-Rf'(R)]-3H\dot{R}f''(R)}\,.
\end{equation}

We note that equation of state parameter of the curvature fluid is
not a constant; it varies actually with redshift. To proceed
further, in which follows we consider the Hu-Sawicki $f(R)$ model
\citep{Hu07,Mar09} given by
\begin{equation}
f(R)=R-m^{2}\,\frac{c_{1}(\frac{R}{m^{2}})^{n}}{c_{2}(\frac{R}{m^{2}})^{n}+1},
\end{equation}
where $m^{2}$, $c_{1}$, $c_{2}$ and $n$ are free positive parameters
that can be expressed as functions of density parameters
\citep{Hu07,Mar09}. Here we explore the dependence of these
parameters on density parameters defined in our setup. To do this
end, we follow the procedure presented in Ref. \citep{Hu07}.
Variation of the action (14) with respect to the metric yields the
induced modified Einstein equations on the brane
\begin{equation}
G_{\alpha\beta}=\frac{1}{M_{5}^{6}}{\cal{S}}_{\alpha\beta}-{\cal{E}}_{\alpha\beta}\,,
\end{equation}
where ${\cal{E}}_{\alpha\beta}$, the projection of the bulk Weyl
tensor on the brane is given by
\begin{equation}
{\cal{E}}_{\alpha\beta}=^{(5)}C_{RNS}^{M}n_{M}n^{R}g_{\alpha}^{N}g_{\beta}^{S}
\end{equation}
and ${\cal{S}}_{\alpha\beta}$ as the quadratic energy-momentum
correction into Einstein field equations is defined as follows
\begin{equation}
{\cal{S}}_{\alpha\beta}=-\frac{1}{4}
\tau_{\alpha\mu}\tau^{\mu}_{\beta}+\frac{1}{12}\tau\tau_{\alpha\beta}+
\frac{1}{8}g_{\alpha\beta}\tau_{\mu\nu}\tau^{\mu\nu}-\frac{1}{24}g_{\alpha\beta}\tau^{2}\,.
\end{equation}
$\tau_{\alpha\beta}$ as the effective energy-momentum tensor
localized on the brane is defined as \citep{Noz08a,Ata08,Saa09}
\begin{equation}
\tau_{\alpha\beta}=-m_{p}^{2}f'(R)G_{\alpha\beta}+\frac{m_{p}^{2}}{2}\Big[f(R)-R
f'(R)\Big]g_{\alpha\beta}+T_{\alpha\beta}+m_{p}^{2}\Big[\nabla_{\alpha}\nabla_{\beta}f'(R)-g_{\alpha\beta}\Box
f'(R)\Big]\,.
\end{equation}
The trace of Eq. (21), which can be interpreted as the equation of
motion for $f'(R)$\,, is obtained as
\begin{equation}
R=\frac{11}{6M_{5}^{6}}\tau^{2}\,.
\end{equation}
$\tau$, the trace of the effective energy-momentum tensor localized
on the brane is expressed as
\begin{equation}
\tau=m_{p}^{2}\Big[2f(R)-Rf'(R)-3\Box f'(R)\Big]-\rho_{m}\,,
\end{equation}
To highlight the DGP character of this generalized setup, we express
the results in terms of the DGP crossover scale defined as
$r_{c}=\frac{m_{p}^{2}}{2M_{5}^{3}}$. So the equation of motion for
$f'(R)$ is rewritten as follows

$$\frac{11}{6}r_{c}^{2}\Bigg (\Big[2f(R)-Rf'(R)\Big]^{2}+9\Big(\Box
f'(R)\Big)^{2}+6Rf'(R)\Box f'(R)-12f(R)\Box f'(R)\Bigg)$$
\begin{equation}
+\frac{11}{3}\frac{r_{c}}{M_{5}^{3}}\Big[Rf'(R)-2f(R)+3\Box
f'(R)\Big]\rho_{m}+\frac{11}{6M_{5}^{6}}\rho_{m}^{2}-R=0\,,
\end{equation}
In the next stage, we solve this equation for $\Box f'(R)$ to obtain
the following solution
\begin{equation}
\Box
f'(R)=-\Bigg[\frac{1}{2}\Big(6Rf'(R)-12f(R)\Big)+\frac{3\rho_{m}}{M_{5}^{3}\,r_{c}}\Bigg]\pm
\sqrt{\Bigg[\frac{1}{2}\Big(6Rf'(R)-12f(R)\Big)+\frac{3\rho_{m}}{M_{5}^{3}\,r_{c}}\Bigg]^{2}-\Theta}\,,
\end{equation}
where $\Theta$ is defined as
\begin{equation}
\Theta=\Bigg[\Big(2f(R)-Rf'(R)\Big)-\frac{1}{M_{5}^{3}\,r_{c}}\rho_{m}\Bigg]^{2}-\frac{6}{11r_{c}^{2}}R\,.
\end{equation}
Now we introduce an effective potential $V_{eff}$, which satisfies
the following equation
\begin{equation}
\Box f'(R)=\frac{\partial V_{eff}}{\partial f'(R)}\,,
\end{equation}
This effective potential has an extremum at $\Theta=0$\,
\begin{equation}
\Big[2f(R)-Rf'(R)\Big]-\frac{1}{M_{5}^{3}\,r_{c}}\rho_{m}=\pm\frac{1}{r_{c}}\sqrt{\frac{6}{11}R}\,,
\end{equation}
In the high-curvature regime, where $f'(R)\simeq 1$ and
$\frac{f(R)}{R}\simeq 1$\,, we recover the standard DGP result (one
can compare this result with corresponding result in Ref.
\citep{Hu07} to see the differences in this extended braneworld
scenario)
\begin{equation}
R\pm\frac{1}{r_{c}}\sqrt{\frac{6}{11}R}=\frac{2}{m_{p}^{2}}\,\rho_{m}\,.
\end{equation}
The negative and positive sign in this equation is corresponding to
the DGP self-accelerating and normal branch respectively. In which
follows, we adopt the positive sign corresponding to the normal
branch of the scenario. To investigate the expansion history of the
universe in our model, we restrict ourselves to those values of the
model parameters that yield expansion histories which are
observationally viable. We note that the Hu-Sawicki $f(R)$ function,
introduced in Ref. \citep{Hu07,Mar09}, was interpreted as a
cosmological constant in the high-curvature regime. The motivation
for that interpretation was to obtain a $\Lambda$CDM behavior in the
high curvature (in comparison with $m^{2}$) regime. Here we would
like to investigate $f(R)$ models that mimic the phantom-like
behavior on the brane in the mentioned regime. As we have pointed
out previously, the phantom-like behavior can be realized from the
dynamical screening of the brane cosmological constant. In this
respect, we apply the same strategy to our model, so that the second
term in the Hu-Sawichi $f(R)$ function ( that is, second term in the
right hand side of equation (20)) mimics the role of an effective
cosmological constant on the DGP brane. Then this term will be
screened by $\frac{H}{\overline{r}_{c}}$ term in the late time (see
the normal
branch of Eq. (16)).\\

In the case in which $R\gg m^{2}$\,, one can approximate Eq. (20) as
follows
\begin{equation}
\lim_{m^{2}/R \rightarrow 0}f(R)\approx
R-\frac{c_{1}}{c_{2}}m^{2}+\frac{c_{1}}{c_{2}^{2}}m^{2}\Big(\frac{R}{m^{2}}\Big)^{n}.
\end{equation}
During the late-time acceleration epoch, $f'_{0}(R)\simeq 1$ or
equivalently $R_{0}\gg m^{2}$ and we can apply the above
approximation. Also the curvature field is always near the minimum
of the effective potential. So, based on Eq. (31), we have
\begin{equation}
R+
\frac{1}{r_{c}}\sqrt{\frac{6}{11}R}=\frac{2}{m_{p}^{2}}\,\rho_{m}+2\frac{c_{1}}{c_{2}}m^{2}\,.
\end{equation}
Since $R$ in the $f(R)$ function is induced Ricci scalar on the
brane, we except crossover scale to affect on the constant
parameters $c_{1}$\,, $c_{2}$ and $m^{2}$. In Ref. \citep{Hu07} they
obtained $3m^{2}\equiv R_{c}=\frac{\rho_{_0m}}{m_{p}^{2}}$ that
$\rho_{_0m}$ is the present value of the matter density, But in our
setup the present value of the matter density (see Eq. (32)) is
given by
\begin{equation}
R_{c}+
\frac{\sqrt{R_{c}}}{r_{c}}\sqrt{\frac{6}{11}}=\frac{2}{m_{p}^{2}}\,\rho_{_0m}\,.
\end{equation}
If we solve this equation for $R_{c}$, we find
\begin{equation}
3m^{2}\equiv
R_{c}=1.1\Omega_{r_{c}}+6\Omega_{m}\pm\sqrt{0.55\Omega_{r_{c}}\Big(0.55\Omega_{r_{c}}+6\Omega_{m}\Big)}\,.
\end{equation}
Therefore, the DGP character of this extended modified gravity
scenario is addressed through $m^{2}$. As we have argued, at the
curvatures high compared with $m^{2}$, the second term on the right
hand side of equation (20) mimics the role of an effective
cosmological constant on the brane. In this respect, the second term
in the right hand side of equation (33) also mimics the role of a
cosmological constant on the brane in the high curvature regime.
With this motivation, we find
\begin{equation}
\frac{c_{1}}{c_{2}}\approx
\frac{18\Omega_{\Lambda}}{1.1\Omega_{r_{c}}+6\Omega_{m}\pm\sqrt{0.55\Omega_{r_{c}}\Big(0.55\Omega_{r_{c}}+6\Omega_{m}\Big)}}\,.
\end{equation}
There is also a relation for $\frac{c_{1}}{c_{2}^{2}}$ as follows
\begin{equation}
\frac{c_{1}}{c_{2}^{2}}=\frac{1-f_{0}'(R)}{n}\bigg(\frac{R_{0}}{m^{2}}\bigg)^{n+1}\,,
\end{equation}
where $\frac{R_{0}}{m^{2}}$ in our setup can be calculated as
follows: firstly, by using Eqs. (34) and (37), we find
\begin{equation} R+
\frac{\sqrt{\frac{6}{11}R}}{r_{c}}=\frac{2}{m_{p}^{2}}\,\rho_{_0m}a^{-3}+12\Omega_{\Lambda}\,,
\end{equation}
where $\rho_{_0m}$ can be omitted through Eq. (26) to obtain
\begin{equation} R+
\frac{\sqrt{\frac{6}{11}R}}{r_{c}}=\bigg(3m^{2}+\frac{m}{r_{c}}\sqrt{\frac{18}{11}}\bigg)a^{-3}+12\Omega_{\Lambda}\,.
\end{equation}
Finally, if we solve this equation for $\sqrt{R}$\,, we find the
following relation for $\frac{R_{0}}{m^{2}}$
\begin{equation}
\frac{R_{0}}{m^{2}}\approx
\Bigg(-\frac{3\Omega_{r_{c}}}{4m}+\Bigg[\Big(\frac{3\sqrt{\Omega_{r_{c}}}}{4m}\Big)^{2}+3\Big(1+
\frac{4\Omega_{\Lambda}}{m^{2}}\Big)+
\frac{5\sqrt{\Omega_{r_{c}}}}{2m}\Bigg]^{1/2}\Bigg)^{2}\,.
\end{equation}
where $m$ is given by Eq. (36). Note that we have set $H_{0}$ and
$a(t_{0})$ equal to unity. These relations tell us that the free
parameters of this model are $n$, $\Omega_{m}$, $\Omega_{r_{c}}$ and
$f_{0}'(R)$, whereas the latter one is constrained by Solar-System
tests. In fact, experimental data show that $f'(R)-1~ <10^{-6}$\,,
when $f'(R)$ is parameterized to be exactly \,$1$ \,in the far past.
To analyze the behavior of $w_{curv}$\,, we need to specify an
ansatz for the scale factor. Here we use the following form
\begin{equation}
a(t)=(t^{2}+\frac{t_{0}}{1-\nu})^{\frac{1}{1-\nu}}
\end{equation}
$\nu\neq 1$ is a free parameter \citep{Cai07}. By noting that the
Ricci scalar is $R=6(\frac{\ddot{a}}{a}+(\frac{\dot{a}}{a})^{2})$,
one can express the function $f(R)$ of equation (20) in terms of the
redshift $z$. Figure $1$ shows the variation of the effective
equation of state parameter versus the redshift. As we see in this
figure, in this class of models the \textit{curvature fluid} has an
effective phantom equation of state, $w_{curv} < -1$ at high
redshifts and then approaches the phantom divide ($w_{curv} = -1$)
at a redshift that decreases by decreasing $n$. The main point here
is that a modified induced gravity of the Hu-Sawicki type in DGP
framework, gives a phantom effective equation of state parameter for
all values of $n$. Note that all of these models reach
asymptotically to the de Sitter phase ($w_{curv}=-1$). As we will
show via a dynamical system approach, this de Sitter phase is a
stable phase.\\

Now using equation (13) we plot the luminosity distance versus the
redshift in this setup. The result is shown in figure $3$ in
comparison with other alternative models. Up to the parameter values
adopted here, the $\Lambda$DGP model has the potential to give a
better fit with $\Lambda$CDM than the $f(R)$-DGP model.

As we have mentioned previously, $w_{curv}$ varies with redshift and
is less than $-1$ which indicates that the curvature fluid plays the
role of a phantom scalar field in the equivalent scalar-tensor
theory. As we will show in the next section, in a dynamical system
approach the accelerating phase of the universe ($q<0$) for this
DGP-inspired $f(R)$ model is stable if we consider the curvature
fluid to be equivalent to a phantom scalar field ( this means that
we set $w_{curv}<-1$ ), while if we set $-1<w_{curv}<-\frac{1}{3}$
(equivalent to a quintessence scalar field ), we find that current
acceleration in the mentioned universe is a transient phenomenon.


\subsection{The phase space of $f(R)$-DGP models}

To investigate stability of the solutions presented in the previous
subsection, here we express the cosmological equations of $f(R)$-DGP
scenario in the form of an autonomous dynamical system. For this
purpose, we define the following normalized expansion variables
\begin{equation}
s=\frac{\sqrt{\Omega_{m}}}{a^{3/2}E}\,,\quad\quad
p=\frac{\sqrt{\Omega_{curv}}}{a^{3(1+w_{curv})/2}E}\,,\quad\quad
u=\frac{\sqrt{\Omega_{r_c}}}{E}\,.
\end{equation}
In this way, equation (16) with minus sign and in a dimensionless
form is written as follows
\begin{equation}
1+2u=s^{2}+p^{2}\,.
\end{equation}
This constrain means that the phase space of this scenario in the
($s$ - $p$) plane is outside of a circle with radius $1$ which is
defined as $s^{2}+p^{2}\geq 1$\,. The autonomous system is obtained
as follows
\begin{equation}
s'=\frac{3s[s^{2}+(1+2w_{curv})p^{2}-1]}{2(s^{2}+p^{2}+1)}\,,
\end{equation}

\begin{equation}
p'=\frac{3p[2s^{2}+(1+w_{curv})(p^{2}-s^{2}-1)]}{2(s^{2}+p^{2}+1)}\,.
\end{equation}
To achieve the critical points of this system one should set $s'=0$
and $p'=0$\,. Note that the critical points and their stability
depend on the value of $w_{curv}$\,. Here we investigate the
stability of critical points in two different subspaces of the model
parameter space where EoS of the curvature fluid has either a
phantom or a quintessence character. In table 1, we see that the
current accelerating phase of the universe expansion is unstable if
the curvature fluid is considered to be a quintessence scalar field
($-1<w_{curv}$) and stable if it is considered to be a phantom field
($w_{curv}<-1$) in the equivalent scalar-tensor theory. This result
is compatible with evolution of the deceleration parameter versus
the redshift as has been shown in figure $4$. When $z\rightarrow
-1$, the accelerating phase (with $q<0$) is stable for
$w_{curv}<-1$. It is necessary to mention that whenever
$w_{curv}=-1$, the phase space is 1D (here the curvature fluid plays
the role of a cosmological constant, the same as $\Lambda DGP$
model. For more details see Ref. \citep{Chi06}). Figure 2 shows the
phase space trajectories of the model.




\subsection{$F(R,\varphi)$-DGP model}

Now we extend the previous model to an even more general case that
the modified induced gravity is non-minimally coupled to a canonical
scalar field on the brane. This extension allows us to see the role
played by the non-minimal coupling of gravity and scalar degrees of
freedom in the cosmological dynamics on the brane. We consider a
general coupling between gravity and scalar degrees of freedom on
the brane \citep[see for
instance][]{Noz08a,Ata08,Saa09,Noz08b,Far00,Noz07a,Bar08,Bam09}. In
fact, inclusion of this field brings the theory to realize a smooth
crossing of the phantom divide line in a fascinating manner
\footnote{ We note that the normal branch of the pure DGP scenario (
which has the capability to describe the phantom like effect),
cannot realize crossing of the phantom divide line without
introducing a quintessence scalar field (a canonical field) on the
brane. Introduction of a quintessence field on the brane, brings the
theory to realize this interesting feature \citep{Chi06}. For
$F(R,\varphi$)-DGP scenario it is natural to expect realization of
the this feature due to wider parameter space in this case
\citep{Noz08a,Noz09a}.}.

The action of this general model is given as follows \citep{Noz09b}
\begin{equation}
S=\frac{M_{5}^{3}}{2}\int d^{5}x\sqrt{-g} {\cal{R}} + \int
d^{4}x\sqrt{-q}\Big(\frac{m_{p}^{2}}{2}
F(R,\phi)-\frac{1}{2}q^{\mu\nu}\nabla_{\mu}\phi\nabla_{\nu}
\phi-V(\phi)+M_{5}^{3}\overline{K}+{\cal{L}}_{m}\Big),
\end{equation}
where the first term shows the usual Einstein-Hilbert action in the
5D bulk. The second term on the right hand side is a generalization
of the Einstein-Hilbert action induced on the brane. This is an
extension of the scalar-tensor theories in one side and a
generalization of $f(R)$-gravity on the other side. $\overline{K}$
is the trace of the mean extrinsic curvature on the brane in the
higher dimensional bulk, corresponding to the York-Gibbons-Hawking
boundary term. We call this model as $F(R,\phi)$-DGP scenario. Note
that from the above action, the crossover scale takes the following
form \citep{Noz09a}
\begin{equation}
l_{F}=\frac{m_{p}^{2}}{2M_{5}^{3}}F'(R,\varphi)=
r_{c}\,\,F'(R,\varphi)\,,
\end{equation}
where as usual $r_{c}=\frac{m_{p}^{2}}{2M_{5}^{3}}$\,, and a prime
denotes a differentiation with respect to $R$. Since DGP scenario
accounts for embedding of the FRW cosmology at any distance scale
\citep{Noz07b}, we start with the following line-element
\begin{equation}
ds^{2}=q_{\mu\nu}dx^{\mu}dx^{\nu}+b^{2}(y,t)dy^{2}=-n^{2}(y,t)dt^{2}+
a^{2}(y,t)\gamma_{ij}dx^{i}dx^{j}+b^{2}(y,t)dy^{2}\,,
\end{equation}
where $\gamma_{ij}$ is a maximally symmetric $3$-dimensional metric
defined as $\gamma_{ij}=\delta_{ij}+k\frac{x_{i}x_{j}}{1-kr^{2}}$.
Also $k=-1,0,1$ parameterizes the spatial curvature and
$r^2=x_{i}x^{i}$. We choose the gauge $b^{2}(y,t)=1$ in the normal
Gaussian coordinates. The cosmological dynamics on the brane in this
model can be described by the following Friedmann equation
\citep[see][]{Noz08a,Ata08,Saa09,Noz09a}
\begin{equation}
H^{2}+\frac{k}{a^{2}}=\frac{1}{3m_{p}^{2}F'(R,\varphi)}\Bigg[\rho^{tot}+
\varrho\Big(1+\epsilon\sqrt{1+\frac{2}{\varrho}
\Big[\rho^{tot}-\frac{m_{p}^{2}F'(R,\varphi)}{a^{4}}{\cal{E}}_{0}\Big]}\Big)\Bigg]
\end{equation}
where $\epsilon=\pm1$ shows two different embedding of the brane in
the bulk,\, $\varrho\equiv\frac{6M_{5}^{6}}{m_{p}^{2}F'(R,\phi)}$\,
and
\,${\cal{E}}_{0}=3\Big(\frac{\dot{a}^{2}}{n^{2}}-a'^{2}+k\Big)a^{2}$\,
is a constant with respect to $y$ ( with $a'\equiv\frac{da}{dy}$),
see \citep{Dic01} for more detailed discussion on the constancy of
this quantity. Total energy density and pressure are defined as
~$\rho^{(tot)}=\rho_{m}+\rho_{\phi}+\rho^{(curv)}+\rho_{\Lambda}$\,
and\, ~$p^{(tot)}=p_{m}+p_{\phi}+p^{(curv)}+p_{\Lambda}$~
respectively. The ordinary matter on the brane has a perfect fluid
form with energy density $\rho_{m}$ and pressure $p_{m}$, while the
energy density and pressure corresponding to non-minimally coupled
quintessence scalar field and also those related to
\textit{curvature fluid} are given respectively as follows
\begin{equation}
\rho_{\varphi}=\bigg[\frac{1}{2}\dot{\phi}^{2}+n^{2}V(\phi)-
6\frac{dF}{d\phi}H\dot{\phi}\bigg]_{y=0},
\end{equation}

\begin{equation}
p_{\phi}=\bigg[\frac{1}{2n^{2}}\dot{\phi}^{2}-V(\phi)+\frac{2}{n^{2}}
\frac{dF}{d\phi}(\ddot{\phi}-\frac{\dot{n}}{n}\dot{\phi})+
4\frac{dF}{d\phi}\frac{H}{n^{2}}\dot{\phi}+\frac{2}{n^{2}}
\frac{d^{2}F}{d\phi^{2}}\dot{\phi}^{2}\bigg]_{y=0}\,,
\end{equation}

\begin{equation}
\rho^{(curv)}=m_{p}^{2}\bigg(\frac{1}{2} \bigg[F(R,\phi)-R
F'(R,\phi)\bigg] -3\dot{R}H F''(R,\phi) \bigg)\,,
\end{equation}
\begin{equation}
p^{(curv)}=m_{p}^{2}\bigg({2\dot{R}H F''(R,\phi)+\ddot{R}F''(R,\phi)
+\dot{R}^{2}F'''(R,\phi)-\frac{1}{2}\Big[ F(R,\phi)-R
F'(R,\phi)\Big]}\bigg).
\end{equation}
Also by definition $\rho_{\Lambda}\equiv\frac{\Lambda}{8\pi G}$.
Ricci scalar on the brane is given by
$$R=3\frac{k}{a^2}+\frac{1}{n^{2}}\bigg[6\frac{\ddot{a}}{a}+
6\Big(\frac{\dot{a}}{a}\Big)^{2}-6\frac{\dot{a}}{a}\frac{\dot{n}}{n}\bigg].$$
We set $n(0,t)=1$ where $y=0$ is chosen to be the location of the
brane. By neglecting the dark radiation term in equation (23), we
find
\begin{equation}
H^{2}=\frac{8\pi
G}{3}(\rho_{m}+\rho_{\varphi}+\rho^{(curv)})+\frac{\Lambda}{3}+\frac{1}{2l_{F}^{2}}+
\epsilon\sqrt{\frac{1}{4l_{F}^{4}}
+\frac{1}{2l_{F}^{2}}\Big[\frac{8\pi
G}{3}(\rho_{m}+\rho_{\varphi}+\rho^{(curv)})+\frac{\Lambda}{3}\Big]}\,,
\end{equation}
where $G=G_{eff}\equiv\frac{1}{8\pi m_{p}^{2}F'(R,\varphi)}$. By
adopting the negative sign, we have
\begin{equation}
H^{2}=\frac{8\pi
G}{3}(\rho_{m}+\rho_{\varphi}+\rho^{(curv)})+\frac{\Lambda}{3}-\frac{H}{r_{c}F'(R,\varphi)}\,.
\end{equation}
Comparing this equation with the following Friedmann equation
\begin{equation}
H^{2}=\frac{8\pi
G}{3}(\rho_{m}+\rho_{\varphi}+\rho^{(curv)})+\rho_{DE}^{eff}\,,
\end{equation}
we are led to conclude that the screening effect on the cosmological
constant is modified by $F'(R,\varphi)$ as follows
\begin{equation}
\frac{8\pi
G}{3}\rho_{DE}^{eff}=\frac{\Lambda}{3}-\frac{H}{r_{c}F'(R,\varphi)}\,.
\end{equation}
To see how this model works, we consider an explicit form of
$F(R,\varphi)$ as
$F(R,\varphi)=\frac{1}{2}(1-\xi\phi^{2})[R-(1-n)\zeta^{2}(R/\zeta^{2})^{n}]$
where $\zeta$ is a suitably chosen parameter \citep[see for
instance][]{Cap03,Sot10,Noj03,Noj04,Noj05,Noj06,Noj07,Noj08,Bam08,Car04,Ame07,Sta07}
and \citep{Cap08,Cap10}. For a spatially flat FRW geometry, the
Ricci scalar is given by $R=6\frac{\ddot{a}}{a}+
6(\frac{\dot{a}}{a})^{2}.$  One can investigate variation of the
effective dark energy density with respect to parameters $\xi$, $t$
and $n$ to see the status of phantom mimicry in this model. As has
been shown in Ref. \citep{Noz09a}, this model accounts for, and
modifies the phantom like behavior on the brane. Using the
conservation equation
$\dot{\rho}_{eff}+3H(1+\omega_{eff})\rho_{eff}=0$, fulfillment of
the condition $w_{eff}<-1$\, requires the constraint
$\frac{\dot{H}}{H}<\frac{\dot{F}'(R,\phi)}{F'(R,\phi)}$
\citep{Noz09a}. In this situation, this model has the potential to
realize the phantom-like behavior and smooth crossing of the phantom
divide line by the effective equation of state parameter. Now, in
order to compare the $F(R,\varphi)$-DGP model with $\Lambda$CDM and
also with other alternative models, we study expansion histories of
these models based on the variation of their luminosity distances
versus the redshift. The evolution of the cosmic expansion in the
normal branch of this F(R,$\varphi$)-DGP setup is given by
$$\frac{H^{2}(z)}{H_{0}^{2}}=\Omega_{m}(1+z)^{3}+\Omega_{\varphi}(1+z)^{3(1+\omega_{\varphi})}
+\Omega_{curv}(1+z)^{3(1+\omega_{curv})}+\Omega_{\Lambda}+2\Omega_{r_{c}}$$
\begin{equation}
-2\sqrt{\Omega_{r_{c}}}\sqrt{\Omega_{m}(1+z)^{3}+\Omega_{\varphi}(1+z)^{3(1+\omega_{\varphi})}
+\Omega_{curv}(1+z)^{3(1+\omega_{curv})}+\Omega_{\Lambda}+\Omega_{r_{c}}}\,\,.
\end{equation}
Note that by definition,
$\Omega_{r_{c}}=\frac{1}{4r_{c}^{2}H_{0}^{2}F'(R,\varphi)}$ and we
use the normalization $F'(R,\varphi)|_{(z=0)}\simeq 1$. This model
has the disadvantage that it contains more free parameters than
$\Lambda$CDM and even $\Lambda$DGP. However, this larger parameter
space provides, at least theoretically,  more facilities to realize
some interesting features. The parameter space of this model can be
constraint at $z=0$ so that
\begin{equation}
1=\bigg(\sqrt{\Omega_{m}+\Omega_{\varphi}+\Omega_{curv}+\Omega_{\Lambda}
+\Omega_{r_{c}}}-\sqrt{\Omega_{r_{c}}}\bigg)^{2}.
\end{equation}
Since $\Omega_{r_{c}}$\,  and \,
$\Omega_{m}+\Omega_{\varphi}+\Omega_{curv}+\Omega_{\Lambda}
+\Omega_{r_{c}}$ \, should be positive, there are two possibilities
for constrains on the parameters of this model as follows
\citep{Sah03,Sah04,Sht00,Lue04}
\begin{equation}
\Omega_{m}+\Omega_{\varphi}+\Omega_{curv}+\Omega_{\Lambda}
-2\sqrt{\Omega_{r_{c}}}=1.
\end{equation}
\begin{equation}
\Omega_{m}+\Omega_{\varphi}+\Omega_{curv}+\Omega_{\Lambda}
+2\sqrt{\Omega_{r_{c}}}=1.
\end{equation}
In our forthcoming analysis, we consider the first one of these
constraints. Finally, using equation (13), we compare the luminosity
distance - redshift relation of this model with other proposed
scenarios. The results are shown in figure $3$. In comparison with
$\Lambda$DGP and $f(R)$-DGP, the $F(R,\varphi)$-DGP scenario has the
potential to have a better fit with $\Lambda$CDM. The parameters
adopted in this model are $\Omega_{m}=0.27$\,,
$\Omega_{r_{c}}=0.01$\,, $\Omega_{\varphi}=0.1$\,,
$\Omega_{curv}=0.13$ and $\Omega_{\Lambda}=0.7$.\\

\subsection{$F(G,\varphi)$-DGP model}

In the $F(G,\varphi)$-DGP setup, the curvature corrections are taken
into account via incorporation of the Gauss-Bonnet invariant term in
the brane part of the action. The Gauss-Bonnet invariant should be
considered essentially in the bulk action. But, if there is a
coupling between scalar degrees of freedom on the brane and the
mentioned invariant term, then it is necessary to consider its
contribution in the field equations on the brane too. This is the
main reason for incorporating scalar field  $\varphi$ in the
$F(G,\varphi)$ setup. We start with the action of this scenario as
follows \citep{Noz09,Noj07,Bam10}
\begin{equation}
{\cal{S}}=\frac{M_{5}^{3}}{2}\int d^{5}x\sqrt{-g} {\cal{R}} +\int
d^{4}x\sqrt{-q}\bigg[\frac{m_{p}^{2}}{2}R+M_{5}^{3}
\overline{K}+F(G,\phi)+{\cal{L}}_{m}\bigg].
\end{equation}
The \emph{Scalar-Gauss-Bonnet} term in this action is defined as
\citep{Noz09}
$$F(G,\phi)\equiv-\frac{1}{2}\partial_{\mu}\phi\partial^{\mu}\phi-V(\phi)+f(\phi)G(R).$$
By definition, the Gauss-Bonnet invariant $G(R)$ is given by
\begin{equation}
G(R)=R^2-4R_{\mu\nu}R^{\mu\nu}+R_{\mu\nu\alpha\beta}R^{\mu\nu\alpha\beta}.
\end{equation}
In order to discuss about cosmological dynamics on the brane, one
can achieve the generalized Friedmann equation as follows
\citep{Noz09}
\begin{equation}
H^{2}+\frac{k}{a^2}=\frac{\rho_{m}+\rho^{(GB)}}{3m_{p}^{2}}
+\frac{1}{2r_{c}^{2}}+\varepsilon\sqrt{\frac{1}{4r_{c}^{4}}+\frac{1}{r_{c}^{2}}
\Big(\frac{\rho_{m}+\rho^{GB}}{3m_{p}^{2}}\Big)}.
\end{equation}
The energy density corresponding to the Gauss-Bonnet term is defined
as
\begin{equation}
\rho^{(GB)}\equiv\frac{1}{2}\dot{\phi}^{2}+V(\phi)-24\dot{\phi}f'(\phi)H^3,
\end{equation}
and the corresponding pressure is defined as
\begin{equation}
p^{(GB)}=\frac{1}{2}\dot{\phi}^{2}-V(\phi)+8\frac{\partial}{\partial
t}\Big(H^2\dot{f}\Big)+16\dot{\phi}f'(\phi)H^3,
\end{equation}
where a dot marks differentiation with respect to the cosmic time.
One can express the Friedmann equation (65) for the normal branch of
this DGP-inspired scenario and for an spatially flat brane in a
dimensionless form as follows
$$\frac{ H^{2}(z)}{
 H_{0}^{2}}=\Omega_{m}(1+z)^{3}+\Omega_{GB}(1+z)^{3(1+\omega_{GB})}+2\Omega_{r_{c}}$$
\begin{equation}
 - 2\sqrt{\Omega_{r_{c}}}\sqrt{\Omega_{m}(1+z)^{3}+\Omega_{GB}(1+z)^{3(1+\omega_{GB})}+
 \Omega_{r_{c}}}
\end{equation}
where by definition $\Omega_{GB}=\frac{8\pi
G}{3H_{0}^{2}}\rho_{0}^{(GB)}$. At the red-shift $z=0$, equation
(68) can be expressed as
\begin{equation}
1=\bigg(\sqrt{\Omega_{m}+\Omega_{GB}+\Omega_{r_{c}}}-\sqrt{\Omega_{r_{c}}}\bigg)^{2}.
\end{equation}
By choosing the positive sign of the square root, this constrain
equation yields
\begin{equation}
\Omega_{m}+\Omega_{GB}-2\sqrt{\Omega_{r_{c}}}=1.
\end{equation}
Note that the general relativistic limit can be recovered if we set
$\Omega_{r_{c}}=0$ (or $M_{5}=0$). In this case equation (70)
implies $\Omega_{m}+\Omega_{GB}=1$. Finally, the luminosity
distance-redshift of this model is compared with other proposed
alternatives in figure $3$.

\section{Some other probes}
To have more complete comparison between the proposed models, in
this section we study some other features of these models. The
current universe is accelerating and this can be realized by
investigating the deceleration parameter. The deceleration parameter
$q=-\frac{\ddot{a}}{\dot{a}^{2}}a$ can be expressed as
\begin{equation}
q(z)=\frac{H'(z)}{H(z)}(1+z)-1\,,
\end{equation}
where a prime marks differentiation with respect to $z$. We study
variation of $q$ versus the redshift in each of the previously
proposed models. For this purpose, we use $H(z)$ as given by
equations (9),\, (12),\, (18),\, (59) and (68). Figure $4$ compares
the deceleration parameter in each of the proposed scenarios. All of
these scenarios explain the late-time cosmic speed-up in their
normal DGP branches, but the redshift at which transition to the
accelerated phase occurs are different in these scenarios. The
$\Lambda$DGP model transits to the accelerated phase at $z\simeq0.8$
much sooner than other scenarios. The $F(R,\varphi)$-DGP model
transits to the accelerated phase at $z\simeq0.4$ much later than
other alternatives. We note however that the exact value of the
transit redshift depends on the choice of the parameters in each
model. Within the parameters values adopted here, the result are
shown in figure $4$. As another important result, we note that
late-time acceleration in the $f(R)$-DGP universe is a stable
phenomenon since we considered the curvature fluid to be a phantom
scalar field. Also the acceleration phase is a transient phenomenon
for $F(G,\varphi)$-DGP model if we set $w_{\varphi}>-1$. Our
inspection shows that cosmological dynamics in the normal branch of
$f(R)$-DGP scenario with a quintessence curvature fluid is very
similar to cosmological dynamics in the normal branch of
$F(G,\varphi)$-DGP scenario for $w_{GB}>-1$.

The next probe is the age of the universe in each of these
scenarios. The age of the universe at a given cosmological red-shift
is given as follow
\begin{equation}
t(z)=\int_{z}^{\infty} \frac{dz'}{(1+z')H(z')}\,,
\end{equation}

where $H(z)$ is given in each of the proposed scenarios. In figure
$5$ the age of the universe is compared in five alternative
scenarios. Within the parameter values adopted here, the age of the
universe in $f(R)$-DGP is larger than $\Lambda$CDM, but this age in
$F(G,\varphi)$-DGP is smaller than $\Lambda$CDM one.

\section{Summary}
In this paper, observational status of the $\Lambda$CDM\,,
$\Lambda$DGP and some extended phantom-like cosmologies as
alternatives for dark energy are studied. We compared these models
with each other by comparing variation of their luminosity distances
versus the redshift and also the age of the universe in each of
these scenarios. The common feature of the mentioned DGP-inspired
scenarios is possibility of realization of the phantom-like behavior
in their normal DGP branches. In the extended scenarios, we focused
on the role played by new ingredients such as the curvature effect,
modification of the induced gravity on the brane and even the
existence of a quintessence field non-minimally coupled to modified
induced gravity on the brane. Since all of our analysis are
preformed on the normal DGP branch of each scenario, there is no
ghost instabilities in these setups. As an example and to have an
intuition about stability of the solutions, we analyzed stability of
cosmological phases of $f(R)$-DGP scenario within a dynamical system
approach. We have analyzed the phase space of the normal branch of
this model where curvature fluid plays the role of a phantom or
quintessence scalar field in the equivalent scalar-tensor theory. We
have shown that the matter dominated phase of this model is a
repeller, unstable point. However, there is a de Sitter phase which
is an attractor point if curvature fluid plays the role of a phantom
scalar field in the equivalent scalar-tensor theory. For a
quintessence field plying the role of the curvature fluid, the de
Sitter phase is an unstable, saddle point. Up to the parameters
values adopted in this paper, $F(R,\varphi)$-DGP scenario mimics the
behavior of the $\Lambda$CDM scenario more than other proposed
models. As another important outcome, all of these scenarios explain
the late-time cosmic speed-up in their normal DGP branches, but the
redshift at which transition to the accelerating phase occurs are
different in these scenarios. For instance, the $\Lambda$DGP model
transits to the accelerating phase at $z\simeq 0.8$, much sooner
than other scenarios. However, the $F(R,\varphi)$-DGP model transits
to the accelerated phase at $z\simeq 0.4$, much later than other
alternatives. Nevertheless, the exact value of the transit redshift
depends on the choice of the parameters in each model. Finally, we
compared the age of the universe in each of the proposed scenarios.
Within the parameter values adopted in this paper, the age of the
universe in the $f(R)$-DGP model is larger than $\Lambda$CDM, but
this age in $F(G,\varphi)$-DGP is smaller than $\Lambda$CDM one.


\begin{figure}[htp]
\begin{center}\includegraphics{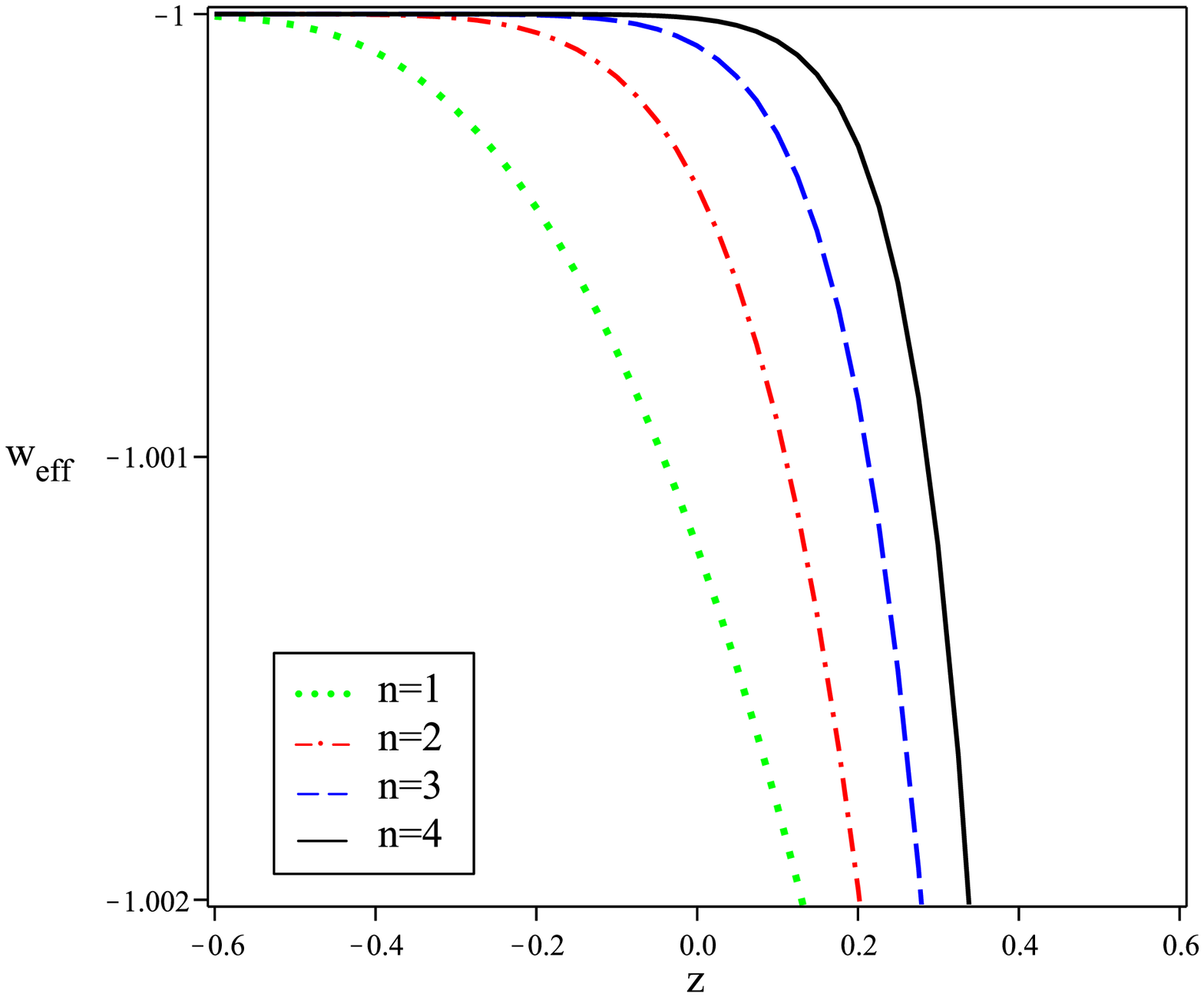} \vspace{7cm}
\end{center}
 \caption{\small {$w_{curv}$ versus the redshift for a DGP-inspired $f(R)$
 model. The $f(R)$ is chosen to be the Hu-Sawicki model.
The selected free parameters in plotting this figure are chosen to
be
$\Omega_{m}=0.27$\,,$\Omega_{\Lambda}=0.93$\,,$\Omega_{r_{c}}=0.01$
and $f_{0}'(R)-1=10^{-6}$\,. As this figure shows, in this class of
models the \textit{curvature fluid} has an effective phantom
equation of state with $w_{curv} < -1$ at high redshifts. This
effective equation of state parameter approaches the phantom divide
line ($w_{curv}= -1$) at a redshift that decreases by decreasing the
value of $n$. }}
\end{figure}

\begin{figure}[htp]
\begin{center}\includegraphics{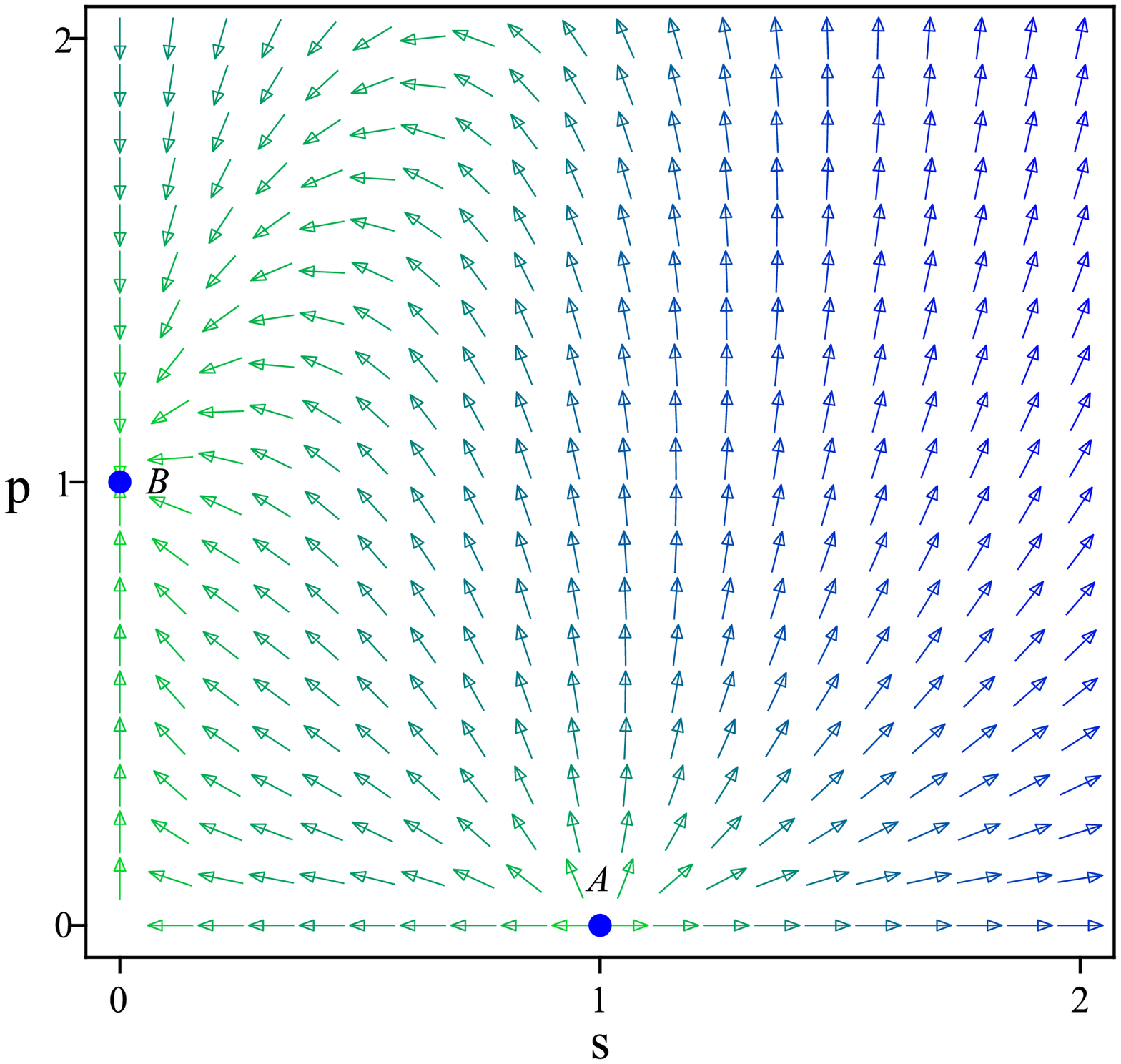} \vspace{3cm}\includegraphics{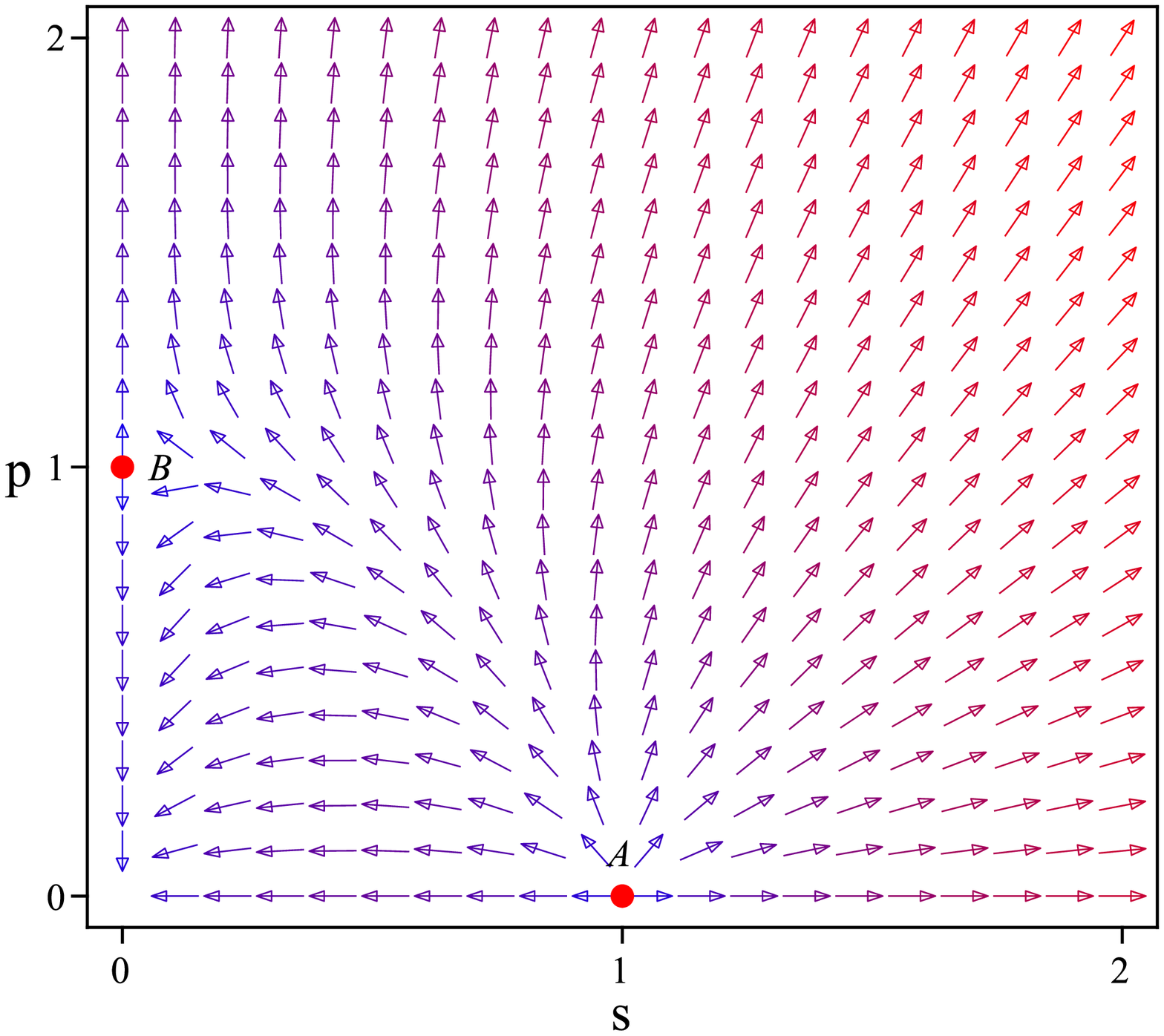} \vspace{3cm}
\end{center}
 \caption{\small {The phase space of the normal branch of DGP inspired $f(R)$ models in which
 curvature fluid plays the role of a phantom scalar field (left) and
 the quintessence scalar field (right) in the equivalent scalar-tensor theories.
 Point $A$ in these two cases is the matter dominated phase and it is a repeller, unstable point.
 Whereas, point $B$ that represents the de Sitter phase, is an attractor point in the
first case and a saddle point in the second case. }}
\end{figure}


\begin{figure}[htp]
\begin{center}\includegraphics{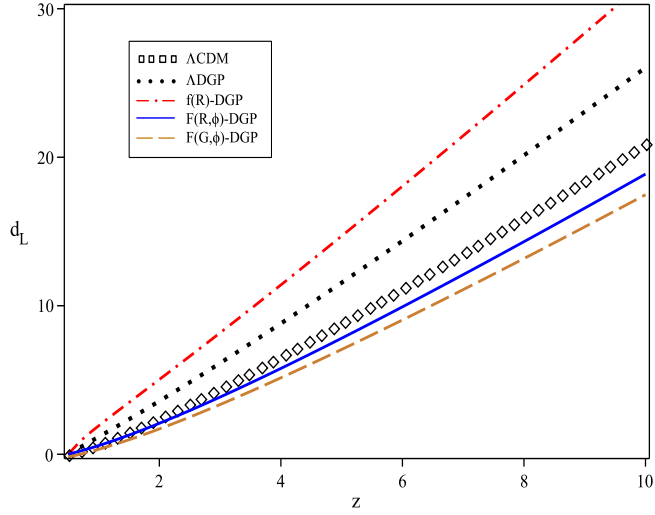} \vspace{6.5cm}
\end{center}
 \caption{\small {Luminosity distance versus the red-shift for five
 alternative scenarios proposed in this paper. In this figure we have set
 $(\Omega_{m},\Omega_{r_{c}})=(0.27,10^{-2})$ for all of the models
 and $(\Omega_{\varphi},\Omega_{\Lambda},\Omega_{curv})=(0.1,0.7,0.13)$ for the $F(R,\varphi)$-DGP setup,
  $\Omega_{\Lambda}=0.93$ for the $\Lambda$DGP,
  $\Omega_{curv}=0.93$ for the $f(R)$-DGP setup and
  $\Omega_{GB}=0.93$ .
 We calculated these quantities using the constraint
 equations attributed to each model. Up to the parameters values
 adopted here, $F(R,\varphi)$-DGP scenario is closer to the
 $\Lambda$CDM than other proposed models.
 We note however that this result depends strongly on the adapted parameters values.}}
\end{figure}

\begin{figure}[htp]
\begin{center}\includegraphics{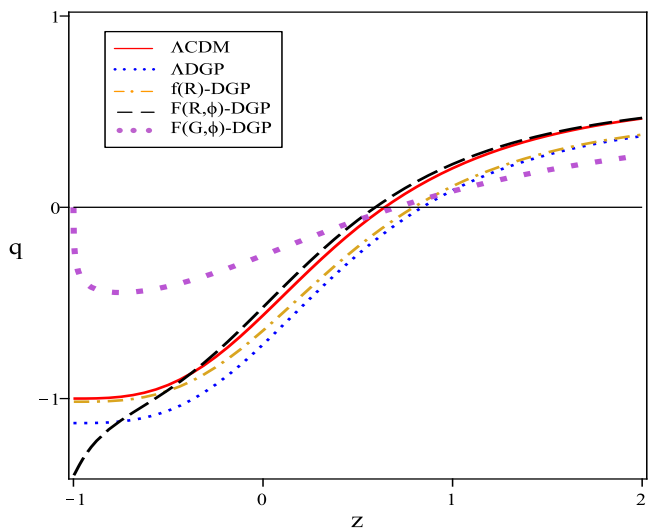} \vspace{7cm}
\end{center}
 \caption{\small {The deceleration parameter $q(z)$ versus the red-shift for
 five alternative scenarios discussed in this paper. The model parameters are
 the same as adopted in figure $3$ and $w_{curv}$ here is adopted quintessence
 like in the left and phantom like in the right. Note that current acceleration in
 the $f(R)$-DGP universe in which \textit{curvature fluid} play role of a quintessence
is a transient and unstable phenomenon but if it act as a phantom
field, it result in a stable accelerated phase for the universe. }}
\end{figure}

\begin{figure}[htp]
\begin{center}\includegraphics{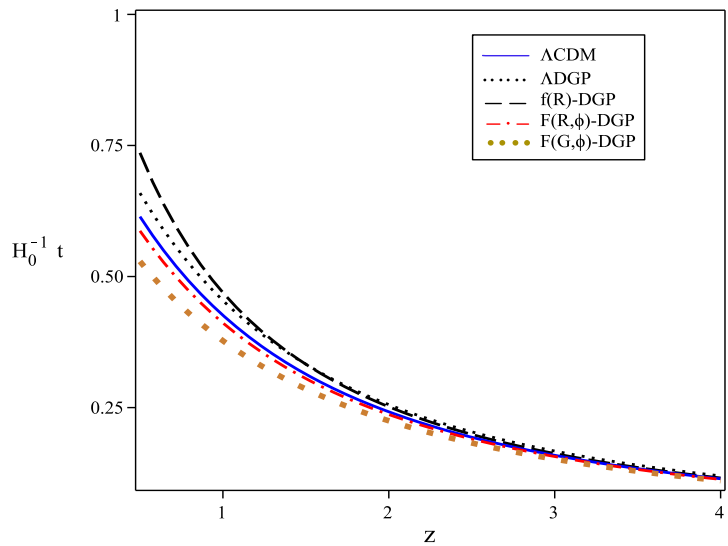} \vspace{8cm}
\end{center}
 \caption{\small {The age of the universe (in units of the inverse Hubble parameter) versus the red-shift
for five alternative scenarios discussed in this paper. The model
parameters are the same as adopted in figure 3. With proposed
parameter values, the $f(R)$-DGP model has the largest and the
$F(G,\varphi)$-DGP has the smallest age. }}
\end{figure}







\clearpage

\begin{table}
\begin{center}
\caption{Eigenvalues and the stability properties of the critical
points.\label{tbl-1}}\vspace{0.5 cm}
\begin{tabular}{ccccc}
\tableline\tableline points & $(s,p)$ & eigenvalues & for $w_{curv}<-1$ & for $(-1<w_{curv}<0)$\\
\tableline
\\
 $A$ &$(1,0)$&$(-\frac{3w_{curv}}{2},\frac{3}{2}) $& unstable & unstable  \\
 \\
$B$ &$(0,1)$&$(\frac{3w_{curv}}{2},\frac{3(1+w_{curv})}{2})$& stable
& unstable \\
\\
\tableline
\end{tabular}
\end{center}
\end{table}


\begin{thebibliography}{}


\bibitem[Perlmutter(1999)]{Per99} Perlmutter, S. {\it et al}
1999, \aj, 517, 565

\bibitem[Riess(1998)]{Rie98} Riess, A. G. 1998, \aj, 116, 1006

\bibitem[Astier et al.(2006)]{Ast06} Astier, P. et al., 2006,
\aa, 447, 31

\bibitem[Wood-Vasey et al.(2007)]{Woo07}
Wood-Vasey, W. M. et al., 2007, \aj, 666, 694

\bibitem[Miller et
al.(1999)]{Mil99} Miller, A. D. et al., 1999, AJL, 524, L1

\bibitem[Hanany(2000)]{Han00} Hanany, S.
2000, AJL, 545, L5

\bibitem[Spergel et al.(2003)]{Spe03}
Spergel, D. N. et al., 2003, AJS, 148, 175

\bibitem[Colless et
al.(2001)]{Col01} Colless, M. et al., 2001, \mnras, 328, 1039

\bibitem[Tegmark et al.(2004)]{Teg04} Tegmark, M et al., 2004,
\prd, 69, 103501

\bibitem[Cole et al.(2005)]{Col05} Cole, S. et
al., 2005, \mnras, 362, 505

\bibitem[Springel et al.(2006)]{Spr06} Springel, V., Frank, C. S., \& White, S. M. D. 2006,
Nature (London), 440, 1137

\bibitem[Sahni \& Starobinsky(2000)]{Sah00} Sahni, V. \& Starobinsky, A. A. 2000, Int. J. Mod.
Phys. D, 9, 373

\bibitem[Padmanabhan(2003)]{Pad03} Padmanabhan,
T. 2003, Phys. Rep, 380, 235

\bibitem[Copeland et al.(2006)]{Cop06} Copeland, E. J., Sami, M., \& Tsujikawa, S. 2006, 15,
1753

\bibitem[Weinberg(1989)]{Wei89} Weinberg, S. 1989, Rev.
Mod. Phys., 61, 1

\bibitem[Carroll(2001)]{Car01} Carroll, S. M.
2001,  Living Rev. Relativity, 4, 1

\bibitem[Caldwell et al.(1999)]{Cal99} Caldwell, R. R., Dave, R., \& Steinhardt, P. 1999,
\prd, 59, 504

\bibitem[Ratra \&
Peebles(1988)]{Rat88} Ratra, B., \& Peebles, P. J. E. 1988, \prd,
37, 3406

\bibitem[Sahni \& Wang(2000)]{Sah00} Sahni, V., \& Wang,
L. 2000, \prd, 62, 103517

\bibitem[Saini et al.(2000)]{Sai00} Saini, T. D., Raychaudhury,
S., Sahni, V., \& Starobinsky, A. A. 2000, \prl, 85, 1162

\bibitem[Brax \& Martin(2000)]{Bra00} Brax, P., Martin, J. 2000,
\prd, 61, 103502

\bibitem[Barreiro et al.(2000)]{Bar00}
Barreiro, T., Copeland, E. J., \& Nunes, N. J. 2000, \prd, 61,
127301

\bibitem[Sahni et al.(2002)]{Sah02} Sahni, V., Sami, M., \&
Souradeep, T. 2002, \prd, 65, 023518

\bibitem[Sami et al.(2003)]{Sam03} Sami, M., Dadhich, N., \& Shiromizu, T. 2003, Phys.
Lett. B, 568, 118

\bibitem[Sami \& Padmanabhan(2003)]{Sam03} Sami, M., \&
Padmanabhan, T. 2003, \prd, 67, 083509

\bibitem[Caldwell(2002)]{Cal02} Caldwell, R. R. 2002, Phys. Lett.
B, 545, 23

\bibitem[Tsujikawa \& Sami(2004)]{Tsu04} Tsujikawa, S., \& Sami,
M. 2004, Phys. Lett. B, 603, 113

\bibitem[Caldwell \& Linder(2005)]{Cal05} Caldwell, R. R., \&
Linder, E. V. 2005, Phys. Rev. Lett., 95, 141301

\bibitem[Cai et al.(2010)] {Cai10} Cai, Y. -F., Saridakis, E. N., Setare, M. R., \& Xia, J.
-Q. 2010, Phys. Reports, 493, 1-60

\bibitem[Moyassari \& Setare (2009)]{Moy09} Moyassari, P. \& Setare,
M. R., 2009, Phys. Lett. B, 674, 237

\bibitem[Kamenshchik et al.(2001)]{Kam01}
Kamenshchik, A., Moschella, U., \& Pasquier, V. 2001, Phys. Lett. B,
511, 265

\bibitem[Dev et al.(2003)]{Dev03} Dev, A., Alcaniz, J. S., \&
Jain, D. 2001, \prd, 67, 023515 37

\bibitem[Amendola et al.(2003)]{Ame03} Amendola, L., Finelli, F., Burigana, C., \& Carturan,
D. 2003, \jcap, 0307, 005

\bibitem[Bertolami et al.(2004)]{Ber04}
Bertolami, O., Sen, A. A., \& Silva, P. T. 2004, \mnras, 353, 329

\bibitem[Biesiada et al.(2005)]{Bie05}
Biesiada, M., Godlowski, W., \& Szydlowski, M. 2005, \aj, 622, 28

\bibitem[Zhang et al.(2006a)]{Zha06a} Zhang, X., Wu, F. -Q., \&
Zhang, J. 2006a, \jcap, 0601, 003

\bibitem[Zhang \& Zhu(2006b)]{Zha06b} Zhang, H., \& Zhu, Z. -H.
2006b, \prd, 73, 043518

\bibitem[Zhang et al.(2009)]{Zha09} Zhang, H., Zhu, Z. -H., \&
Yang, L. 2009, Mod. Phys. Lett. A, 24, 541

\bibitem[Heydari-Fard \&
Sepangi(2008)]{Hey08} Heydari-Fard, M., \& Sepangi, H. R. 2008,
\prd, 78, 064007

\bibitem[Roos(2007)]{Roo07} Roos, M. 2007, [arXiv:0704.0882]

\bibitem[Bouhmadi-L\'{o}pez \& Lazkoz(2007)]{Bou07}
Bouhmadi-L\'{o}pez, M., \& Lazkoz, R. 2007, Phys. Lett. B, 654, 51

\bibitem[Roos(2008a)]{Roo08a} Roos, M. 2008a, [arXiv:0804.3297]

\bibitem[Roos(2008b)]{Roo08b} Roos, M. 2008b, Phys. Lett. B, 666,
420

\bibitem[Setare(2009)]{Set09} Setare, M. R. 2009, Int. J. Mod. Phys. D, 18,
419

\bibitem[Capozziello(2003)]{Cap03} Capozziello, S., V. F.,
Cardone, Carloni, S., \& Troisi, A. 2003, Int. J. Mod. Phys. D, 12,
1969

\bibitem[Sotiriou(2010)]{Sot10} Sotiriou, T. P., \& Faraoni, V.
2010, [ arXiv:/0805.1726]

\bibitem[Nojiri \& Odintsov(2007)]{Noj07} Nojiri, S., \&
Odintsov, S. D. 2007, Int. J. Geom. Meth. Mod. Phys., 4, 115

\bibitem[Nojiri \& Odintsov(2004)]{Noj04} Nojiri, S., \& Odintsov, S. D. 2004, Gen.
Relat. Gravit., 36, 1765

\bibitem[Nojiri \& Odintsov(2003)]{Noj03} Nojiri, S., \&
Odintsov, S. D. 2003, \prd, 68, 123512

\bibitem[Nojiri \& Odintsov(2005)]{Noj05} Nojiri, S., \&
Odintsov, S. D. 2005, Class. Quantum Grav., 22, L35

\bibitem[Nojiri \& Odintsov(2006)]{Noj06} Nojiri, S., \& Odintsov,
S. D. 2006, \prd, 74, 086009

\bibitem[Nojiri \& Odintsov(2008)]{Noj08} Nojiri, S., \&
Odintsov, S. D. 2008, \prd, 78, 046006

\bibitem[Bamba et al.(2008)]{Bam08} Bamba, K., Nojiri, S., \&
Odintsov, S. D. 2008, \jcap, 0810, 045

\bibitem[Carroll et al.(2004)]{Car04} Carroll, S. M., Duvvuri, V., Trodden, M., \& Turner,
M. S. 2004, \prd, 70, 043528

\bibitem[Amendola et al.(2007)]{Ame07} Amendola, L., Polarski,
D., \& Tsujikawa, S. 2007, Phys. Rev. Lett., 98, 131302

\bibitem[Starobinsky(2007)]{Sta07} Starobinsky, A. 2007, JTEP, 86,
157

\bibitem[Nozari(2008a)]{Noz08a} Nozari, K., \&
Pourghassemi, M. 2008a, \jcap, 10, 044

\bibitem[Atazadeh et al(2008)]{Ata08} Atazadeh, K., Farhoudi, M.,
\& Sepangi, H. R. 2008, Phys. Lett. B, 660, 275

\bibitem[Saavedra \& Vasquez(2009)]{Saa09} Saavedra, J., \&
Vasquez, Y. 2009, \jcap, 04, 013

\bibitem[Setare (2008)]{Set08} Setare, M. R. 2008, Int. J. Mod. Phys. D, 17,
2219

\bibitem[Dvali et al.(2000a)]{Dva00a} Dvali, G. R., Gabadadze, G., \&
Porrati, M. 2000a, Phys. Lett. B, 484, 112

\bibitem[Deffayet(2001)]{Def01} Deffayet, C. 2001, Phys. Lett. B, 502, 199

\bibitem[Lue(2006)]{Lue06} Lue, A. 2006, Phys. Rept., 423, 48

\bibitem[Dvali et al.(2000b)]{Dva00b} Dvali, G. R., Gabadadze, G., \& Porrati, M. 2000b, Phys. Lett. B, 485, 208

\bibitem[Dvali \& Gabadadze(2001)]{Dva01} Dvali, G. R., \& Gabadadze, G. 2001, \prd, 63, 065007

\bibitem[Dvali et al.(2002)]{Dva02} Dvali, G. R., Gabadadze, G., Kolanovi$\acute{c}$, M., \& Nitti, F. 2002, \prd, 65, 024031

\bibitem[Melchiorri et al.(2003)]{Mel03} Melchiorri, A., Mersini, L., Odman, C. G., \& Trodden, M. 2003, \prd, 68, 043509

\bibitem[Riess et al.(2004)]{Rie04} Riess, A. G. et al. 2004, \aj, 607, 665

\bibitem[Komatsu et al.(2009)]{Kom09} Komatsu, E. et al. [WMAP Collaboration] 2009, AJS, 180, 330

\bibitem[Sahni \& Shtanov(2003)]{Sah03} Sahni, V., \& Shtanov, Y. 2003, \jcap, 0311, 014 0005193

\bibitem[Sahni et al.(2005)]{Sah05} Sahni, V., Shtanov, Y., \& Viznyuk, A. 2005, \jcap, 0512, 005

\bibitem[Sahni(2004)]{Sah04} Sahni, V. 2004, [arXiv:astro-ph/0502032]

\bibitem[Shtanov(2000)]{Sht00} Shtanov, Y. V. 2000, [arXiv:hep-th/0005193]

\bibitem[Lue \& Starkman(2004)]{Lue04} Lue, A., \& Starkman, G. D. 2004, \prd, 70, 101501

\bibitem[Collins \& Holdom(2000)]{Col00} Collins, H., \& Holdom, B. 2000, \prd, 62, 105009

\bibitem[Lazkoz et al.(2006)]{Laz06} Lazkoz, R., Maartens, R., \& Majerotto, E. 2006, \prd, 74, 083510

\bibitem[Maartens \& Majerotto(2006)]{Maa06} Maartens, R. \& Majerotto, E. 2006, \prd, 74, 023004

\bibitem[Lazkoz \& Majerotto(2007)]{Laz07} Lazkoz, R. \& Majerotto, E. 2006, \jcap, 07, 015

\bibitem[Alam \& Sahni(2006)]{Ala06} Alam, U. \& Sahni, V. 2006, \prd, 73, 084024

\bibitem[Sollerman et al.(2009)]{Sol09} Sollerman, J. et al. 2009, \aj, 703, 1374

\bibitem[Nozari(2009a)]{Noz09a} Nozari, K., \& Kiani, F. 2009a, \jcap, 07, 010

\bibitem[Bouhmadi-Lopez(2009)]{Bou09} Bouhmadi-Lopez, M. 2009, \jcap, 0911, 011

\bibitem[Nozari(2009b)]{Noz09b} Nozari, K., \& Rashidi, N. 2009b, \jcap, 0909, 014

\bibitem[Nozari(2009c)]{Noz09c} Nozari, K., \& Azizi, T. 2009c, Phys. Lett. B, 680, 205

\bibitem[Nozari(2009d)]{Noz09d} Nozari, K., \& Aliopur, N. 2009d, Europhys. Lett., 87, 69001

\bibitem[Atazadeh \& Sepangi(2006)]{Ata06} Atazadeh, K., \& Sepangi, H. R. 2006, Phys. Lett. B, 643, 76

\bibitem[Atazadeh \& Sepangi(2007)]{Ata07} Atazadeh, K., \& Sepangi, H. R. 2007, \jcap, 0709, 020

\bibitem[Atazadeh \& Sepangi(2009)]{Ata09} Atazadeh, K., \& Sepangi, H. R. 2009, \jcap, 01, 006

\bibitem[Hu \& Sawicki(2007)]{Hu07} Hu, W. \& Sawicki, I. 2007, \prd, 76, 064004

\bibitem[Martinelli et al.(2009)]{Mar09} Martinelli, M., Melchiorri, A., \& Amendola, L. 2009, \prd, 79, 123516

\bibitem[Cai et al.(2007)]{Cai07} Cai, Y. -Fu., Qiu, T., Piao, Y. -S., \& Zhang, X. 2007, JHEP, 071, 0710

\bibitem[Faraoni(2000)]{Far00} Faraoni V. 2000, \prd, 62, 023504

\bibitem[Nozari(2007a)]{Noz07a} Nozari, K. 2007a, \jcap, 0709, 003

\bibitem[Nozari(2008b)]{Noz08b} Nozari, K., \& Fazlpour, B. 2008b, \jcap, 0806, 032

\bibitem[Barenboim \& Lykken(2008)]{Bar08} Barenboim, G., \& Lykken, J. 2008, \jcap, 0803, 017

\bibitem[Bamba et al.(2009)]{Bam09} Bamba, K., Geng, C. -Q., Nojiri, S., \& Odintsov, S. D. 2009, \prd, 79, 083014

\bibitem[Chimento et al.(2006)]{Chi06} Chimento, L. P., Lazkoz, R., Maartens, R., \& Quiros, I. 2006, \jcap, 0609, 004

\bibitem[Nozari(2007b)]{Noz07b} Nozari, K. 2007b, Phys. Lett. B, 652, 159

\bibitem[Dick(2001)]{Dic01} Dick, R. 2001, Class. Quant. Grav., 18, R1

\bibitem[Capozziello et al.(2008)]{Cap08} Capozziello, S., Cardone, V. F., \& Salzano, V. 2008, \prd, 78, 063504

\bibitem[Capozziello et al.(2010)]{Cap10} Capozziello, S.,
Laurentis, M., \& Faraoni, V. 2010, [ arXiv:0909.4672]

\bibitem[Nozari et al.(2009)]{Noz09} Nozari, K., Azizi, T., \& Setare, M. R. 2009, \jcap, 010, 022

\bibitem[Nojiri et al.(2007)]{Noj07} Nojiri, S., Odintsov, S. D., \& Tretyakov, P. V. 2007, Phys. Lett. B, 651, 224

\bibitem[Bamba et al.(2010)]{Bam10} Bamba, K., Odintsov, S. D., Sebastiani, L.,
\& Zerbini, S. 2010, [arXiv:0911.4390]

\end{thebibliography}
\end{document}